\documentclass[acp, manuscript]{copernicus}

\usepackage{tikz}
\usetikzlibrary{patterns}
\usetikzlibrary{shapes}
\usetikzlibrary{decorations.text}
\usetikzlibrary{shapes.multipart}
\usetikzlibrary{arrows,shapes.geometric,positioning}
\usetikzlibrary{shapes,positioning,decorations.pathmorphing}
\usetikzlibrary{calc}
\usepackage{booktabs}
\usetikzlibrary{backgrounds}
\usepackage{varwidth}
\usetikzlibrary{fit}

\usepackage{bm,bbm}
\usepackage{amsmath,amsfonts,amssymb,mathrsfs}
\usepackage{latexsym,euscript,textcomp}
\usepackage{color,graphics,epsf,dcolumn}
\usepackage{epstopdf,ulem}
\usepackage{blindtext}
\usepackage{threeparttable}

\usepackage{enumitem}

\newcommand{\beq}{\begin{equation}}
\newcommand{\eeq}{\end{equation}}

\allowdisplaybreaks

\begin{document}
\nolinenumbers

\title{How transpiration by forests and other vegetation determines alternate moisture regimes}


\Author[1,2]{Anastassia M.}{Makarieva}
\Author[1]{Andrei V.}{Nefiodov}
\Author[3]{Antonio Donato}{Nobre}
\Author[4]{Ugo}{Bardi}
\Author[5,6,7]{Douglas}{Sheil}
\Author[8]{Mara}{Baudena}
\Author[9]{Scott R.}{Saleska}
\Author[10]{Anja}{Rammig}

\affil[1]{Theoretical Physics Division, Petersburg Nuclear Physics Institute, 188300 Gatchina, St.~Petersburg, Russia}
\affil[2]{Institute for Advanced Study, Technical University of Munich, Lichtenbergstrasse 2~a, 85748 Garching, Germany}
\affil[3]{Centro de Ci\^{e}ncia do Sistema Terrestre INPE, S\~{a}o Jos\'{e} dos Campos, 12227-010 S\~{a}o Paulo, Brazil}
\affil[4]{Department of Chemistry, University of Florence, Italy}
\affil[5]{Forest Ecology and Forest Management Group, Wageningen University \& Research, PO Box 47, 6700 AA, Wageningen, The Netherlands}
\affil[6]{Center for International Forestry Research (CIFOR), Kota Bogor, Jawa Barat, 16115, Indonesia}
\affil[7]{Faculty of Environmental Sciences and Natural Resource Management, Norwegian University of Life Sciences, \AA s, Norway}
\affil[8]{National Research Council of Italy, Institute of Atmospheric Sciences and Climate (CNR-ISAC), Turin, Italy}
\affil[9]{Department of Ecology and Evolutionary Biology, University of Arizona, Tucson, 85721, Arizona, USA}
\affil[10]{Technical University of Munich, School of Life Sciences, Hans-Carl-von-Carlowitz-Platz 2, 
85354 Freising, Germany}


\runningtitle{Transpiration control of forest ecosystem{\textquoteright}s two moisture regimes}

\runningauthor{Makarieva et al.}

\correspondence{A. D. Nobre (anobre27@gmail.com), A. M. Makarieva (ammakarieva@gmail.com)}

\received{}
\pubdiscuss{} 
\revised{}
\accepted{}
\published{}


\firstpage{1}

\maketitle

\begin{abstract}
The terrestrial water cycle links the soil and atmosphere moisture reservoirs through four fluxes: precipitation, evaporation, runoff and atmospheric moisture convergence. Each of
these fluxes is essential for human and ecosystem well-being. However, predicting how the water cycle responds to changes in vegetation cover, remains a challenge \citep{lawrence15,ellison2017,tewierik2021}.  Recently, rainfall was shown to decrease disproportionally with reduced 
forest transpiration following deforestation \citep{baudena21}.  Here, combining  these findings with the law of matter conservation, we show that in a sufficiently wet atmosphere forest transpiration  can control atmospheric moisture convergence such that increased transpiration enhances atmospheric moisture import.  Conversely, in a drier atmosphere increased transpiration reduces atmospheric moisture convergence and runoff.  This previously unrecognized dichotomy can explain the seemingly random observations of runoff and soil moisture sometimes increasing and sometimes reducing in response to re-greening \citep[e.g.,][]{zheng2021}. Evaluating the transition between the two regimes is crucial both for characterizing the risk posed by deforestation as well as for motivating and guiding global ecosystem restoration.
\end{abstract}

\introduction[\color{black}Precipitation, column moisture and mass conservation]  

\label{intro}

\begin{figure*}[!tb]
\begin{minipage}[p]{1\textwidth}
\centering\includegraphics[width=0.77\textwidth,angle=0,clip]{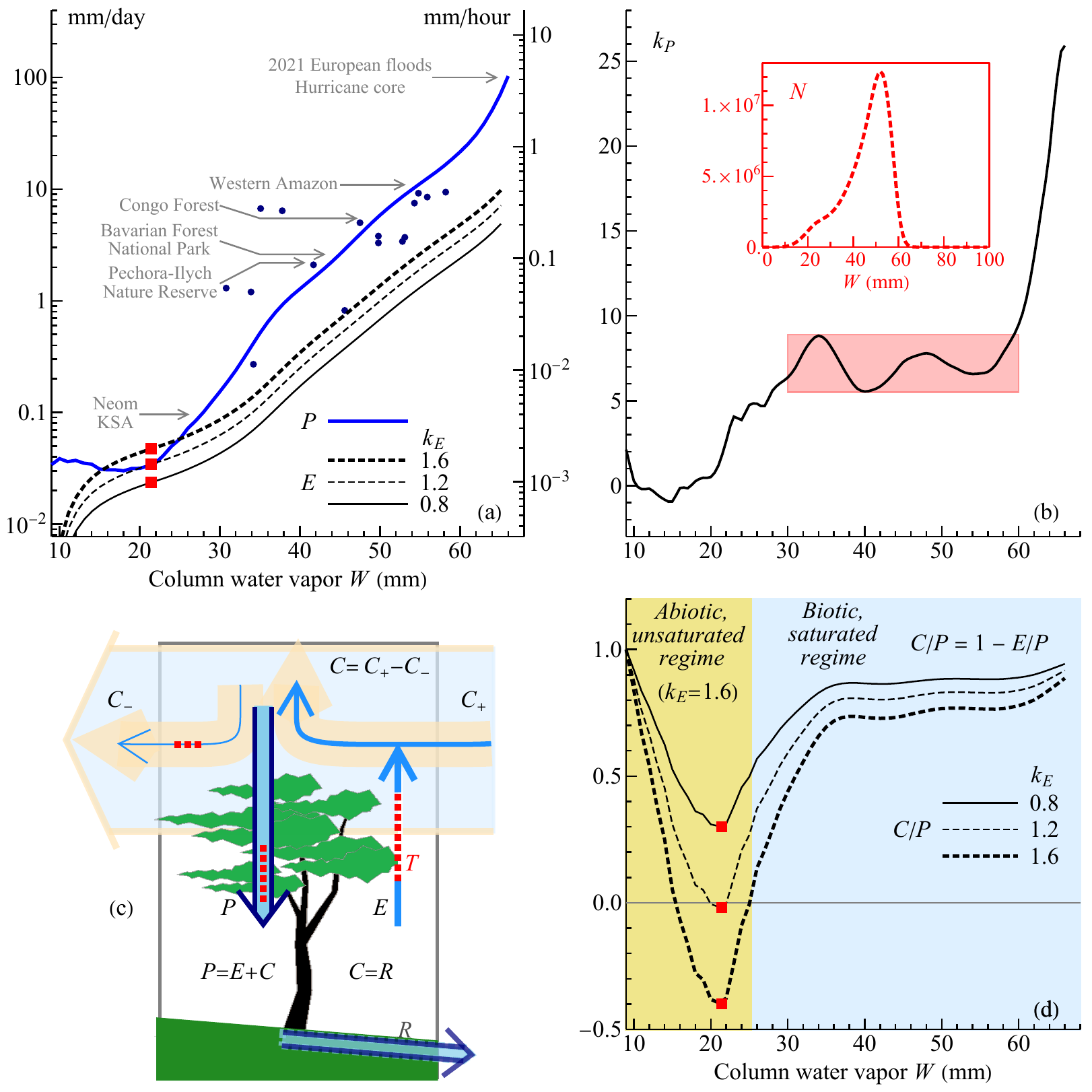}
\end{minipage}
\caption{
Atmospheric moisture budget terms $P$, $E$ and $C$ as related to column water vapor content $W$.  Hourly precipitation data \citep[from][re-averaged arithmetically]{baudena21} give the blue curve in (a),  which we then integrate according to Eq.~\eqref{E} to obtain $E$ (black curves in (a)) and $k_P(W)$ (the black curve in (b)). 
The inset in (b) shows the number of data points $N(W)$ for each 1 mm bin \citep{baudena21}.
Constrained by the mass balance Eq.~\eqref{mb} illustrated in (c), we derive moisture convergence $C$ (d) from $P(W)$ and $E(W)$.   Circles in (a) represent daily averaged $P$ and $W$ 
for several meteostations in Brazil \citep[][Table~2, columns 4 and 11]{jhm14}; arrows indicate characteristic precipitation rates in different locations in the world (see text).
In (c), the steady-state water balance is shown for a forest location that receives moisture solely from the ocean  \citep[thin blue arrows indicate moisture inflow $C_{+}$ and outflow $C_{-}$, cf.][their Fig.~12]{eltahir94}. A certain part of the recycled transpired moisture (red dashed lines) flows to the forest further inland, the rest precipitates locally. Thick yellowish arrows indicate the ascending and descending air motions that generate precipitation and are responsible for local moisture convergence $C > 0$ (and steady-state runoff $R = C$). In (a) and (d), red squares indicate the point where  the moisture convergence (and runoff) are minimal, but begin to grow with increasing $E$ at larger $W$.
}
\label{fig1}
\end{figure*}

Forests play an important role for atmospheric moisture generation and transport through transpiration and atmospheric moisture convergence. Deforestation and reforestation can therefore strongly alter the hydrological cycle depending on the prevailing climate regime in the region. Based on data for a tropical island, \citet{holloway10} showed that the rainfall probability rises sharply with increasing column water vapor $W$ \citep[see also][]{yano22}. Using the radiosonde data for several meteostations in Brazil (Fig.~\ref{fig1}a), \citet{jhm14} concluded that, in the Amazon forest, a small relative increment of $W$ should lead to a larger relative increment in rainfall probability. \citet{baudena21} found that in the relatively flat part of the Amazon including the Amazon-Cerrado transition zone (0$-$18$^{\rm o}$S and 65$-$50$^{\rm o}$W), the mean hourly precipitation $P$ (ERA5 data for 2002-2013, binned for each 1 mm of $W$) depends non-linearily on $W$ (Fig.~\ref{fig1}a)
\beq\label{P}
\frac{dP}{P} = k_P (W)\frac{dW}{W}, 
\eeq
where $k_P \gg 1$ for $W > 30$~mm, while $k_P \simeq 0$ for $10~{\rm mm}< W < 20$~mm (Fig.~\ref{fig1}b).
(We excluded observations at $W \le 8$~mm and $W \ge 67$~mm from our consideration (all bins with less than 20,000 data points, or about $0.1\%$ of all observations, see inset in Fig.~\ref{fig1}b), since $k_P$ displayed an erratic behavior apparently due to the relatively small number of rainfall events at these extreme $W$ values.)

Interestingly, relationship \eqref{P} established for local hourly rainfall for a specific study area in the Amazon region \citep{baudena21} encompasses  characteristic precipitation rates over a broad range of spatial and temporal scales, from annual
precipitation in deserts, temperate and tropical forests, to hurricane and flood-causing rainfall \citep[Fig.~\ref{fig1}a, data from][]{mgl13,ar17,Smirnova2017,beudert2018,almazroui2020,kreienkamp2021}. 
 
Evapotranspiration $E$ adds moisture to the atmosphere, so we expect that $W$ grows with $E$.
We can write such a dependence as follows:
\beq\label{E}
\frac{dE}{P} = k_E(W)\frac{dW}{W},
\eeq
where $k_E > 0$. 
In the Amazon basin, with mean transpiration $T = 45$~mm~month$^{-1}$  \citep{staal2018}
and mean annual rainfall of $P =2200$~mm~year$^{-1}$ \citep{ma06}, a loss of evapotranspiration
due to the loss of transpiration from the entire basin is equal to $\Delta E/P = -T/P = -0.25$, i.e.
a decrease of 25\% (in proportion to precipitation). 
According to \citet{baudena21}, zeroing transpiration in a region reduces $W$ in a given location by the fraction of water vapor originating from
transpiration in the considered region. In the Amazon basin, the fraction of water vapor 
originating from the Amazon transpiration is equal to $0.32$ \citep{staal2018}.
With $\Delta W/W = -0.32$ and $\Delta E/P = -T/P = -0.25$, we have $k_E = 0.8$.

For a steady-state atmospheric column, the vertically integrated continuity equation representing the water vapor convergence over the whole atmospheric column as equal to the difference of precipitation and evaporation fluxes (see scheme in Fig.~\ref{fig1}c), reads:
\beq\label{mb}
C {\color{black} \equiv -\int \limits_{0}^{\infty} \mathrm{div} (\rho_{v} \mathbf{u}) dz} = {\color{black} - \int \limits_{0}^{\infty} \dot{\rho}_{v} dz \equiv} P - E,  
\eeq
where {\color{black} $\mathbf{u}$ is the air velocity vector, $\rho_{v}$ is water vapor denstity, $\dot{\rho}_{v}$ is the mass source/sink and} $C$ is water vapor convergence (net flux of water vapor through the column surface). The mass balance \eqref{mb} can be applied both locally and on a regional scale. It is valid when  the rate of change in moisture content is small compared to precipitation ($dW/dt \ll  P$ ), i.e.,  on a time scale $\tau \lesssim \tau_a$, where $\tau_a \equiv W/P$ (a few days) is the time scale of atmospheric moisture turnover via precipitation. 
{\color{black}
In a system of units where liquid water density  $\rho_{l} = 10^3$~kg m$^{-3}$ is set to unity, column water vapor 
$W=\int \limits_{0}^{\infty} \rho_{v} dz$ has the units of length (mm), while $C$, $P$ and $E$ are measured in units of length per unit of time (for example, mm~day$^{-1}$). 
}

\section{\color{black}Ecosystem{\textquoteright}s two moisture regimes}

Using the dependence $P(W)$, Eq.~(\ref{P})  we can integrate Eq.~(\ref{E}) assuming $E(W_{\rm min}) = 0$ for minimal water vapor content $W_{\rm min} = 9$~mm. 
Then we can find $C(W)$ combining Eqs.~\eqref{P} and \eqref{E} 
\beq\label{C}
dC = dP- dE = (k_P - k_E)dE.
\eeq 
This reveals two regimes, for high and low $W$. In the drier regime with $k_P < k_E$, moisture convergence {\it declines} with increasing evapotranspiration and moisture content,  while precipitation remains relatively constant.  In the wetter regime with $k_P > k_E$, moisture convergence {\it increases} together with evapotranspiration, moisture content and precipitation (Fig.~\ref{fig1}d). (If the soil moisture content is steady, runoff  $R$ is equal to moisture convergence $C$ and thus behaves similarly.)

The quantitative details of the $P(W)$, $E(W)$ and $C(W)$ dependencies will be specific to each region, season and ecosystem type and depend on the spatial and temporal scale of observations \citep[Fig.~\ref{fig1}a, see also][]{peters2006}. However, the established pattern is qualitatively robust with respect to different values of $E(W_{\rm min})$ (not shown) and $k_E$ (Fig.~\ref{fig1}d). At higher $k_E$ (i.e., a slower accumulation of atmospheric moisture with growing $E$), there appears an interval of $W$ with negative moisture convergence.  This corresponds to dry conditions when the ecosystem transpires at the expense of previously accumulated soil moisture or at the expense of irrigation.

Equations~(\ref{P})--(\ref{mb}) make it clear that both regimes exist irrespective of specific $k_E$ and $k_P$ values, provided that $k_P < k_E$ at lower $W$ and $k_P > k_E$ 
at higher $W$. The robustness of the pattern has a profound physical meaning related to the saturated state of the atmosphere. Since precipitation requires saturation, at low $W$ far from the saturation,  $k_P$ is low, i.e., an increase in local $W$ does not markedly increase the probability precipitation that may then be determined by non-local weather systems \citep{jhm14}.   Evaporation, on the contrary, is significant at high water vapor deficits and enriches the atmosphere with water vapor, so {\it if $E$ grows}, so will $W$. This dry regime can be characterized as {\textquotedblleft}abiotic{\textquotedblright}, because the ecosystem exploits the geophysical moisture flows and,  at $C < 0$, the previously accumulated water stores (Fig.~\ref{fig1}d).

As the air column approaches saturation at high $W$, precipitation begins to increase markedly with $W$. Evaporation, on the other hand, depends
on the moisture deficit near the surface atmosphere, which is largely decoupled from total water vapor content $W$ \citep[][their Fig.~3e,f]{holloway09}.
(Indeed, evapotranspiration and moisture convergence have distinct physics. Moisture convergence occurs when the air rises and water vapor condenses.  In contrast, evapotranspiration is not explicitly linked to directional air motions; it adds water vapor directly to the atmospheric column facilitated by turbulent diffusion.)
Under nearly saturated conditions, evaporation can only proceed if precipitation depletes moisture from the atmosphere creating a water vapor deficit \citep{murakami2006,murakami2021,jimenez2021}. Therefore, at high $W$ precipitation $P$ and evapotranspiration $E$ should 
grow approximately proportionally to each other. The recycling ratio $E/P$ and the runoff-to-precipitation ratio $R/P = C/P = 1 - E/P$ stabilize at high values of $W$ and then remain approximately constant (Fig.~\ref{fig1}d). In this wet regime all the components of the water cycle should be under biotic control.

The wetter the atmosphere, the stronger the water cycle control and the more resilient the forest: by slightly changing evapotranspiration, it can compensate for unfavorable disturbances of the water cycle \citep{jhm14}. Conversely, in a drier atmosphere with less rainfall, the forest is more vulnerable to external perturbations. This elucidates why the Amazon forest appears to be losing resilience mostly in the regions where the rainfall is relatively low (rather than where rainfall is decreasing) \citep{boulton2022}.

\section{\color{black}Conclusions}

Our findings, that forest transpiration can control and switch atmospheric moisture convergence, corroborate the biotic pump concept \citep{hess07,jhm14}. Independent studies have shown that the Amazon forest transpiration during the late dry season moistens the atmosphere and triggers the wet season and associated ocean-to-land moisture inflow \citep{wright17}. This dry-season transpiration has a phenological and, hence, evolutionary component that encodes the Amazon dry-season greening \citep{saleska2016}. Enhanced forest transpiration preceding the wet season was also observed in Northeast India and Congo basin \citep{pradhan19,worder2021}. Regulation of moisture convergence could explain how forests buffer precipitation extremes across continents \citep{oconnor2021,deoliveira2021}. While over a broad range of $30~{\rm mm} \lesssim W \lesssim 60$~mm  the value of $k_P$ is relatively constant (pink area in Fig.~\ref{fig1}b), it increases sharply at larger $W$. The interval of $W > 60$~mm harbors very high precipitation rates observed under extreme weather conditions (Fig.~\ref{fig1}a).
It remains to be investigated whether/how forest disturbances influence the probability of such extremes.
If natural forest ecosystems have evolved to stabilize and sustain the continental water cycle, their destruction contributes to destabilization and impoverishment of regional water cycles. This contribution is underestimated \citep{sheil19}. Future studies of vegetation cover impacts on atmospheric moisture flows must emphasize the role of natural forests \citep{zemp17b,makarieva20b,leitefilho21,hua2022}.

The existence of the two regimes with a possibly non-steady transition between them have important implications for large-scale afforestation efforts e.g., in China where increasing atmospheric moisture content with progressive re-greening can be in drier areas associated with a decline of runoff and soil moisture content \citep[e.g.,][]{jiang13,zhang2021}. This decline has been interpreted as limiting further ecosystem restoration \citep[][]{feng2016}. However, Fig.~\ref{fig1}d suggests that if re-greening is continued, the ecosystem can pass a tipping point and enter a wetter state when further re-greening will enhance both rainfall, moisture convergence and runoff. Indeed, in the wetter areas in China re-greening did cause an increase in runoff \citep[e.g.,][]{wang2018}. On the Loess Plateau, soil moisture content decreased in the drier, but increased in the wetter, parts of the region \citep[][Fig.~10a]{zheng2021}. Establishing key parameters of the two regimes and assessing the potential transition from the drier to the wetter state (during which the recovering ecosystem might require extra water inputs) can inform and guide afforestation and reforestation strategies, including possible recovery of ecosystem productivity in the driest regions.

\section*{Acknowledgments}
We thank Dr. Arie Staal for his useful comments on an early draft of this work. Work of A.M. Makarieva is funded by the Federal Ministry of Education and Research (BMBF) and the Free State of Bavaria under the Excellence Strategy of the Federal Government and the L\"ander, as well as by the Technical University of Munich -- Institute for Advanced Study.

\bibliographystyle{copernicus}

\end{document}